\documentstyle[12pt,epsf]{article}
\begin{document}
\baselineskip=24pt
\pagestyle{plain}

\begin{center}
{\Large \bf Conical Singular Solutions  \\ 
in (2+1)-Dimensional Gravity \\  
Employing the ADM Canonical Formalism}
\\
\vspace{1cm}
{\large M. Kenmoku \footnote[1]{kenmoku@phys.nara-wu.ac.jp},
S. Uchida \footnote[2]{uchida@physics.fukui-med.ac.jp} 
and T. Matsuyama \footnote[3]{matsuyat@nara-edu.ac.jp}}
\\
\vspace{5mm}
{\small \it \footnotemark[1] 
Department of Physics, Nara Women's University, Nara 630-8506, Japan \\
\footnotemark[2] 
Department of Physics, Fukui Medical University, Fukui 910-1193, Japan \\
\footnotemark[3] 
Department of Physics, Nara University of Education,  
Nara 630-8528, Japan \\}
\end{center}

\vspace{3cm}

\begin{abstract}
Topological solutions in the (2+1)-dimensional 
Einstein theory of gravity are studied 
within the ADM canonical formalism.  
It is found that a conical singularity appears in the closed de Sitter 
universe solution as a topological defect  
in the case of the Einstein theory with a cosmological constant.  
Quantum effects on the conical singularity are studied   
using the de Broglie-Bohm interpretation.  
Finite quantum tunneling 
effects are obtained for the closed de Sitter universe, 
while no quantum effects are obtained for an open 
universe.  
\end{abstract}

\newpage
\section{Introduction}

A quantum theory of gravity is required 
to unify the four fundamental interactions and
to reduce the singular behavior of the early universe 
and of the central region of black holes.  
Superstring theory \cite{Pol}
and M-theory \cite{M} are very promising approaches  
but the analysis of these theories have not yet been completed,  
because non-perturbative effects 
are difficult to evaluate.  
Recently, the interesting  
correspondence between the gravity theory 
in (d+1)-dimensional anti-de Sitter space (AdS) 
and the conformal field theory (CFT) in d-dimensions 
has been elucidated 
using the concept of duality 
in the superstring theory  
\cite{AdS/CFT}. 
The microscopic derivation of the black hole entropy 
has also been studied extensively 
to apply the AdS/CFT correspondence \cite{Strom}. 

In several works of particular interest, 
it has been shown that 
the Einstein theory of gravity 
in (2+1)-dimensional anti-de Sitter space 
is equivalent to the Chern-Simons gauge theory 
\cite{Achucarro86,Witten88},  
while black hole entropy has been studied 
in terms of its asymptotic 
symmetry \cite{Brown86,Carlip95,Ezawa95,Banados96,Yoshida99}. 
The (2+1)-dimensional gravity theory is a simple toy model 
that possesses no local or physical modes 
(like gravitational wave modes), 
but it does possess interesting global and topological modes. 

\begin{figure}
\epsfxsize=11cm \epsfysize=3.5cm
\centerline{\epsfbox{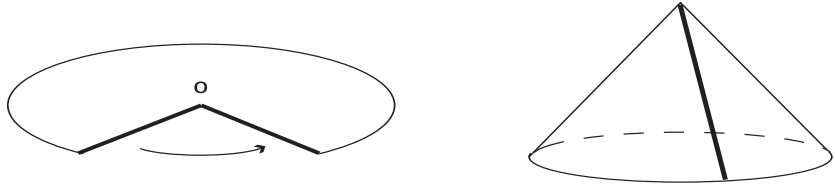}}
\caption{(A) The deficit angle $8\pi m_0G=2\pi m_0$ is shown.  
(B) The conical singularity is formed gluing edges of the deficit angle shown in (A).  } 
\label{fig:conifig2.eps}
\end{figure}

A conical singularity is one of the charasteristic features of 
the (2+1)-dimensional gravity theory \cite{Deser84}.
In the case of a point particle with mass $m_0$ 
located at the origin of the spatial coordinates,  
we consider a static spherically symmetric metric of the form  
\begin{equation}
ds^2=-dt^2+{\rm e}^{\sigma(r)}(dr^2+r^2d\phi^2) \ .
\end{equation}
In this case, the Einstein equation becomes 
\begin{equation}
\bigtriangleup \sigma = m_0 \delta^{(2)}(\vec r) \ ,
\end{equation}
and we obtain its solution as 
\begin{equation}
ds^2=-dt^2+r^{-8m_0 G}(dr^2+r^2d\phi^2) \;,
\end{equation}
where the ranges of the variables are given by 
$ 0\leq r<\infty $ and $\ 0\leq \phi<2 \pi $. 
In order to transform to a flat metric, 
we make a change of variables 
$r\rightarrow \overline{r}$ ,$\phi \rightarrow \overline{\phi}$,
defined by the following: 
\begin{eqnarray} 
\bar r &=& r^{1-4m_0 G}/(1-4m_0 G) \;, \\
\bar \phi &=&(1-4m_0 G)\phi \;, \\
ds^2 &=& -dt^2+\bar d\bar r^2+\bar r^2d\bar \phi^2 \;. 
\end{eqnarray}
The ranges of the variables 
$\overline{r}$ and $\overline{\phi}$ become 
$ 0\leq \bar r <\infty $ and $ 0\leq \bar \phi<2 \pi-8\pi m_0 G $.
Here the deficit angle $8\pi m_0 G$ appears, 
and this leads to a conical singularity (see Fig.\ref{fig:conifig2.eps}). 

In this paper, we study the 
(2+1)-dimensional Einstein gravity 
theory with a cosmological constant 
within the framework of the traditional canonical formalism   
\cite{Dirac,ADM}. 
We quantize the theory in order that the constraint operators 
form a closed algebra. 
We obtain quantum de Sitter universe solutions 
applying the dBB interpretation 
and rederive the known classical  
de Sitter universe solutions 
as the classical limits. 

The dBB interpretation is a natural extension of the WKB approach 
(i.e., the semi-classical method) 
into the full quantum region \cite{dBB}. 
In this interpretation, 
the wave function 
in polar coordinates, $\Psi=\mid \Psi \mid {\rm e}^{i\Theta}$ 
is understood with the following meaning:  
(a) the amplitude {\it squared} $ \mid \Psi\mid^2 $ 
is consedered the probability distribution of metrics,  
and 
(b) the momenta are defined by the gradient of the phase 
according to  
\begin{equation}
\displaystyle
\frac{\partial \Theta}{\partial ({\rm metric})}
=({\rm momentum }) \ \label{dBBequation}.
\end{equation}
We call this the de Broglie-Bohm (dBB) equation. 
The dBB interpretation has 
the following advantageous features 
in the quantum gravity theory:  
(i) A metric including quantum effects can be obtained 
from the dBB equation (\ref{dBBequation});  
(ii) a kind of time is introduced through the dBB equation 
\cite{time}, 
which is lost in the constraint equations
[see Eq. (\ref{constraint}) in Section 2];  
(iii) neither an outside observer nor
a measurement is necessary  
to realize the classical universe from the quantum universe 
asymptotically.  

This paper is organized as follows. 
In Section 2, we review the canonical quantization 
of the Einstein theory of gravity in (2+1) dimensions  
and apply the dBB interpretation to the wave function.  
The notation and our basic idea follow those of 
Kucha\v{r}'s work \cite{Kuchar} and 
our previous paper \cite{Kenmoku2001}.  
In Section 3,  
we study the conical singularity of the closed 
de Sitter universe solution.
This solution is interesting 
with regard to two points:   
(1) the correlation between local objects (black holes)  
and the global object (the de Sitter universe) and 
(2) quantum effects on the systems as implied by the de Broglie-Bohm 
interpretation.   
A summary of the paper is given in Section 4.

\section{Quantum Solutions and the De Broglie-Bohm Interpretation}
We start from the Einstein-Hilbert action with a 
cosmological constant $\lambda$ 
in (2+1)-dimensional spacetime, 
\begin{equation}
I=\frac{1}{4 \pi}
\int d^3x \, \sqrt{-{\rm det} g}\, (R-2\lambda) \;,
\end{equation}
where the gravitational constant is set to $G_{2}={1/4}$.
We write the metric of the general spherically symmetric spacetime 
in the (2+1) decomposition \cite{ADM,Kuchar,Kenmoku2001} as 
\begin{eqnarray}
ds^2=-(N^t)^2dt^2&+&\Lambda^2(dr+N^rdt)^2
+R^2(d\phi + N^{\phi}dt)^2 \nonumber \\
&+&2C(dr+N^rdt)(d\phi + N^{\phi}dt) \ ,
\end{eqnarray}
where the metric is a function of $r$ and $t$. 

The action in the canonical formalism is written in the form  
\begin{equation}
I=\int dt dr (P_{\Lambda} \dot {\Lambda} + 
P_R \dot R +P_C\dot C
-N^tH_t-N^rH_r-N^{\phi}H_{\phi}) \;,
\end{equation}
where
$P_{\Lambda},P_R$ and $P_C$ are the canonical momenta, 
\begin{eqnarray}
P_{\Lambda}&=&\frac{\Lambda R(N^r R'-\dot R)}{N^{t}\sqrt{h}} \ ,\ \\
P_R&=&\frac
{R[C(N^{\phi})'+\Lambda ((\Lambda N^r)'-\dot \Lambda)]}{N^{t}\sqrt{h}} \ ,\ \\
P_C&=&-\frac{(N^r C)'+R^2 (N^{\phi})'-\dot C}{2N^{t}\sqrt{h}} \ , 
\label{momenta}
\end{eqnarray}
where
\begin{eqnarray}
h=R^2\Lambda^2-C^2 \ . 
\end{eqnarray} 
In the above, the dot and prime denote the differentiation 
with respect to 
$t$ and $r$, respectively.  
The coefficients of the auxiliary metrics 
$N^t,N^r$ and $N^{\phi}$ are the Hamiltonian $H_t$ and 
the momentum 
constraint functions $H_r$ and $H_{\phi}$. 
In place of these constraint functions, we consider 
the angular momentum $\hat{J}$ , the mass function $\hat{M}$ 
and one of the momuntum constraint function $\hat{H_{r}}$ 
as the new constraint functions. 
These are given by  
\begin{eqnarray}
\hat J &=& 
-\int dr H_{\phi} 
=
\frac{C}{\Lambda}\hat P_{\Lambda}+R^2\hat P_C \ , 
\label{J} \\
\hat H_r &=& R'\hat P_R-\Lambda(\hat P_{\Lambda})'-C(\hat P_C)' \ ,  \\
\hat M &=& 
-\int dr (\frac{R'}{\bar \Lambda}H_t
+\frac{P_{\Lambda}}{\Lambda}H_r+P_CH_{\phi})
\nonumber \\
&=&
\frac{1}{2} \biggl( \frac{\bar \Lambda}{ \Lambda}
A\hat P_{\Lambda}A^{-1}\frac{\bar \Lambda}{ \Lambda}
\hat P_{\Lambda}-\frac{R'^2}{\Lambda ^2}+ 
1-\lambda R^2+\frac{\hat J^2}{R^2} \biggl) \ \label{M},
\end{eqnarray}
where $\bar \Lambda=\sqrt{\Lambda^2-C^2/R^2}$,  
and the function $A$ in Eq.(\ref{M}) is the ordering factor, 
which is determined 
so as to close the algebra formed by the constraints. 
[Its explicit expression is given in Eq. (\ref{A}).]  
The commutation relations among the new constraint functions 
describing a closed algebra are  
\begin{eqnarray}
\, [\hat J(r), \hat J(r')]&=&0 \;,
\label{JJ} \\ 
\, [\hat J(r), \hat H_r (r') ] &=&i\hat J'(r) \delta(r-r') \;, \\
\, [\hat J(r), \hat M(r')]&=&0 \;, \\
\, [ \hat H_r (r) ,  \hat H_r (r') ]&=&
i( \hat H_r (r)\delta'(r-r') -(r\leftrightarrow r')) \;, \\
\, [\hat H_r (r),\hat M(r')]&=&i\hat M'(r)\delta(r-r') \;, \\
\, [\hat M(r),\hat M(r')]&=&0 \; . 
\label{MM}
\end{eqnarray}

We impose these new constraint functions 
on the wave function $\Psi$  
\begin{eqnarray}
\hat J(r)\Psi=j\Psi \ \  , \ \ 
\hat H_r(r) \Psi=0 \ \ , \ \ 
\hat M(r) \Psi=m\Psi \ \label{constraint},  
\end{eqnarray}
where $j$ and $m$ are constants of integration  
that have the physical meanings of angular momentum and mass. 
 
It is shown that  
the the original constraint functions  
$H_t,H_r$ and $H_{\phi}$ form a closed algebra automatically,   
because 
the new constraint functios 
form a closed algebra expressed by Eqs. (\ref{JJ})-(\ref{MM})
and there is a linear relation between the new and old 
consitraint functions given in Eqs. (\ref{J})-(\ref{M}).   
It is also worthwhile to note that 
the original constraints on the wave function  
are satisfied automatically, if 
the new constraints on the wave function   
expressed by Eq. (\ref{constraint}) are satisfied.

A possible solution of the constraints on the wave function 
given by Eq. (\ref{constraint}) can be written in terms of 
two Hankel functions as 
\begin{eqnarray}
\Psi=\exp{(ij\Phi)}Z^{\nu}[b_1H_{\nu}^{(1)}(Z)+
b_2H_{\nu}^{(2)}(Z)] \ ,
\label{wavefunction}
\end{eqnarray}
where $b_1$ and $b_2$ are constants of integration and  
the variables $\Phi$ and $Z$ are functions of 
the metric defined as 
\begin{eqnarray}
\displaystyle
\Phi&=&\int dr \frac{C(r)}{R(r)^2} \ \ , \ \ 
Z=\int dr\int^{\bar \Lambda}d\bar \Lambda
\sqrt{{\chi}-F_j(R)} \ \ , \\
\chi&=&\frac{R'^2}{\bar \Lambda ^2} \ \ , \ \ 
F_j(R)= 1-2m +\frac{j^2}{4R^2} -\lambda R^2 \ .
\end{eqnarray}
In Eq. (\ref{wavefunction}), the 
arbitrary number $\nu$ denotes the order of the Hankel functions 
and is  
introduced through our choice of 
the ordering factor $A$ appeared in Eq. (\ref{M}) 
(see Ref. \cite{Kenmoku2001} for a detailed explanation):  
\begin{eqnarray}
A&=&A_Z (Z) \  \bar{A}(R,\chi) \ , \nonumber \\
A_Z&=&Z^{2\nu-1} \ , \  \bar A=\sqrt{\chi -F_j} \label{A}\ . 
\end{eqnarray}

We adopt the dBB interpretation in order to 
extract geometrical information from the wave function. 
For this purpose, we should impose the Vilenkin boundary condition  
\cite{Vilenkin88}, which requires the classical expanding universe 
in the large universe limit 
(in our case, this corresponds to the large $Z$ limit). 
Taking into the Vilenkin boudary condition, 
we choose only the Hankel function of the second kind; 
that we set $b_1=0$ in Eq. (\ref{wavefunction}),  because 
the asymptotic behavior of the wave function then becomes 
that of the outgoing wave given by  
\begin{equation}
 H_{\nu}^{(2)} 
\sim  \sqrt{\frac{2}{\pi Z}}\  {\rm e}^{-i(Z-(2\nu +1)\pi/4)}  
\ \ \ {\rm {for \ large}}\ Z\ \ . 
\end{equation}

The dBB equations for the metric are obtained 
from the definition of the dBB equation 
(\ref{dBBequation}) and 
the expression of the momenta (\ref{momenta}) as 
\begin{eqnarray}
\frac{1}{N^t} \biggl( N^r \frac{\partial}{\partial r}R
-\frac{\partial}{\partial t} 
R \biggl) &=& \sqrt{\chi -F_j}
\,\frac{d\Theta}{dZ} \ , 
\label{dbbeq1}
\\
\frac{1}{N^t} \biggl[ \frac{\partial}{\partial r}(\bar \Lambda N^r)-
\frac{\partial}{\partial t} {\bar \Lambda} \biggl]
&=& \frac{\bar \Lambda}{(\partial R/ \partial r)} 
\frac{\partial}{\partial r}\biggl( \sqrt{\chi -F_j} \biggl)
\,\frac{d\Theta}{dZ} \ ,
\label{dbbeq2}
\\
-\frac{R^3}{2{N^t}\bar \Lambda}
\biggl[ 
\frac{\partial}{\partial r}\biggl( \frac{C N^r}{R^2} 
+N^{\phi}\biggl)-\frac{\partial}{\partial t}
{ \biggl( \frac{C}{R^2} \biggl)} \biggl] &=& j \ .  
\label{dbbeq3}
\end{eqnarray}
The classical limit is realized asymptotically,     
\begin{equation}
\displaystyle
\frac{d\Theta}{dZ} =
- \frac{2}{\pi Z \mid H^{(2)}_{\nu}(Z)\mid ^2}  
\rightarrow -1 \; \ \ {\rm{as}}\ Z \rightarrow \infty \ ,
\end{equation}
and the dBB equation approaches the Einstein equation 
in the same limit.

\section{Conical singularity in the De Sitter universe }

\begin{figure}
\epsfxsize=11cm \epsfysize=6cm
\centerline{\epsfbox{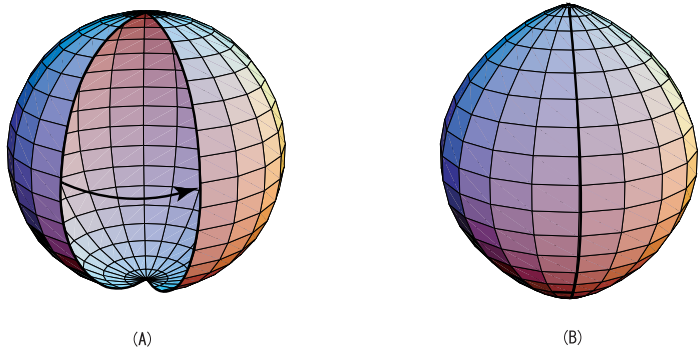}}
\caption{(A) The deficit angle $2\pi(1-\sqrt{1-2m})\sim 2\pi m$ is shown.
(B) The conical singularity in the closed surface is formed gluing edges of the deficit angle shown in (A).}
\label{fig:defsphere.eps}
\end{figure}

In order to obtain the explicit form of the metric, 
we must impose the coordinate condition (i.e. the gauge fixing). 
The proper choice of the coordinate condition for the 
de Sitter universe is 
\begin{equation}
\bar{\Lambda}=
\frac{a(t)}{\sqrt{1-Kr^2}}\ \ \ , \ \ R=b(t)r 
\ \ \ \ {\rm and} \ \ \  j=0  \ , 
\end{equation}
where $K$ is a curvature parameter. 
The cosmological scale factors  
$a(t)$ and $b(t)$ are determined by 
the dBB equations (\ref{dbbeq1})-(\ref{dbbeq3}) as   
\begin{eqnarray}
a(t) &=& {1 \over \sqrt{1-2m} }\, b(t) \;,
\\
\dot a(t) &=& -\sqrt{\lambda a(t)^2-K}\,\frac{d\Theta}{dZ} \ . 
\label{dBBscale}
\end{eqnarray}
In classical limit ($ {d\Theta}/{dZ}\rightarrow -1$), 
the scale factor is obtained as
\begin{equation}
a(t)_{cl}=\left\{ 
\begin{array}{ll}
{\exp}{(\sqrt \lambda (t-t_0))}
&  \ \mbox{for $K=0$  (flat universe) }\\
{(\sqrt{\lambda})}^{-1}\cosh{(\sqrt \lambda (t-t_0))} 
&  \ \mbox{for $K=1$  (closed universe) } \ ,
\end{array}
\right.
\end{equation}
where $t_0$ is a constant of integration, 
which specifies the origin of the time.   
The classical metric becomes 
\begin{equation}
ds^2_{cl}=\left\{
\begin{array}{ll}
-dt^2+ {\exp}{(2\sqrt\lambda (t-t_0))}\ 
(dr^2+r^2 d\bar\phi^2) 
&  \ \mbox{for $K=0$} \\
-dt^2+ {\lambda}^{-1}{\cosh^2(\sqrt\lambda (t-t_0))}
({dr^2}/({1-r^2})+r^2d\bar\phi^2) 
&  \ \mbox{for $K=1$} \ .
\end{array}
\right.
\end{equation}
We note that the range of 
the new angle variable $\bar{\phi}=\sqrt{1-2m}\ \phi$ 
is given by   
\begin{equation}
0\leq\bar\phi<\sqrt{1-2m}\ 2\pi \ ,
\end{equation}
and the deficit angle 
($2\pi(1-\sqrt{1-2m})\approx 2\pi m$) 
leads to the conical singularity 
in the de Sitter universe solution(see Fig.\ref{fig:defsphere.eps}).

\begin{figure}
\epsfxsize=9cm \epsfysize=6cm
\centerline{\epsfbox{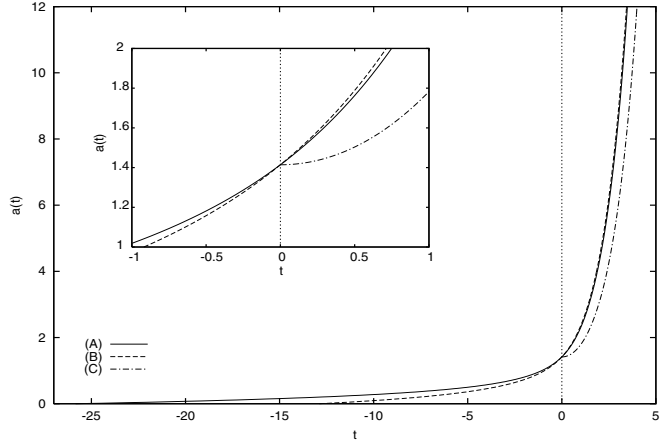}}
\caption{(A) The quantum scale factor 
in the case without a conical singularity ($m= 0$). 
(B) The quantum scale factor in the case with a 
conical singularity ($m\neq 0$). 
(C) The classical scale factor in the case 
with a conical singularity ($m\neq 0$).  
The classically allowed region is $t>0$ and 
the classically forbidden region is $t<0$. 
Each scale factor is normalized such that 
 $a(t=0)=\sqrt{1/\lambda}$.
The parameters are chosen as $\lambda=1/2\ , \ m=1/5$.  }
\label{fig:coniclala1cp.eps}
\end{figure}

Next, we attempt to determine the scale factor of the universe 
including the quantum effect using the dBB equation. 
The explicit form of the dBB equation for the scale factor 
given by Eq. (\ref{dBBscale}) becomes 
\begin{equation}
\dot a(t) =\sqrt{\lambda a(t)^2-K}\, 
\frac{2}{\pi Z \mid H_{\nu}^{(2)}(Z)\mid ^2} \ , 
\label{dota(t)}
\end{equation}
where the variables in the case of de Sitter universe are 
\begin{eqnarray}
Z&=&\int dr (\bar \Lambda \sqrt{\chi-F_j}
-\sqrt{1-2m}\, 
a(t) \ln \mid \frac{\sqrt \chi 
+ \sqrt{\chi -F_j}}{\sqrt{\mid F_j \mid}}\mid ) \ ,
\nonumber \\
\bar\Lambda &=& {a(t)}/{\sqrt{1-Kr^2}} \ ,
\nonumber \\
\chi&=&(1-2m)(1-Kr^2)
\ , \nonumber \\ 
F_j&=&(1-2m)(1-\lambda a(t)^2r^2) \ . \nonumber 
\end{eqnarray}
The quantum effect is included in the factor
 ${2}/({\pi Z \mid H^{(2)}_{\nu}(Z)\mid ^2})$ 
in Eq. (\ref{dota(t)}). 
For the flat de Sitter universe ($K=0$), 
no quantum effect is obtained,  
because $Z$ becomes infinite 
as a result of due to the spatial integration 
over the infinite range of the variable $r$. 
For the closed de Sitter universe ($K=1$), 
a finite quantum effect is obtained 
only in the special case $\nu=1/3$.  
The result of the numerical analysis 
of the dBB equation in this special case
with the scale factor $a(t)$ 
for the closed de Sitter universe given by Eq. (\ref{dota(t)}) 
are presented  
in Fig. \ref{fig:coniclala1cp.eps}. 
These results confirm that 
the quantum scale factors with and without the conical singularity 
approache the
classical scale factor  
asymptotically as 
\begin{eqnarray}
a(t)\mid_{m\ne 0} \approx a(t)\mid_{m= 0} \approx 
 a(t)_{cl} \approx \exp{(\sqrt{\lambda}t)} \ \ 
{\rm{as}}\   t \rightarrow {\rm large} \ . \label{asymptotically}  
\end{eqnarray}

The quantum effects appear remarkably 
near the beginning of the universe 
as the derivation from the classical solution.  
We can show the relations for  
differentiation of the scale factor $a(t)$ 
with respect to $t$     
at the beginning of the classical universe $a(t)=1/\sqrt{\lambda}$ 
\begin{eqnarray}
\dot{a}(t)\mid_{m\ne 0}
&=&(1-2m)^{-1/6}\frac{3^{4/3}\lambda^{1/6}}{4\pi}\Gamma{(2/3)} 
=(1-2m)^{-1/6}\dot{a}(t)\mid_{m= 0}\ \ \nonumber \\ 
&>& \dot{a}(t)\mid_{m= 0}\ \ \nonumber \\ 
&>& \dot{a}(t)\mid_{cl}=0 .  
\end{eqnarray} 
These relations 
together with the asymptotic relations 
for the scale factor given by 
Eq. (\ref{asymptotically})  
imply the following hierarchy relations for $\dot a(t)$ 
 at a given of $a(t)$ 
\begin{eqnarray}
\dot a(t)\mid_{m\neq 0}\ 
> \ \dot a(t)\mid_{m = 0}\ 
> \ \dot a(t)_{cl} 
\ \ . \label{hierarchy}
\end{eqnarray}
The validity of these relations can be seen directly 
from Fig. \ref{fig:coniclala1cp.eps}. 
From these result given by Eq. (\ref{hierarchy}), 
we confirm that the quantum effect 
becomes large as the mass parameter becomes large.

\section{Summary} 
We have studied the (2+1)-dimensional Einstein theory 
of gravity  
with a cosmological constant within the framework of the ADM 
canonical formalism employing the de Broglie-Bohm interpretation. 
We examined the conical singularity in the de Sitter universe 
as a topological effect on the geometry. 
The classical limit is realized for this theory 
in the large universe limit, $a(t)\rightarrow \infty$. 
No quantum effect is obtained for the flat de Sitter universe 
solution, because in this case 
the volume of the universe  
and the value of $Z$ both become infinite.   
A finite quantum effect is obtained for the closed 
de Sitter universe solution with the parameter choice $\nu=1/3$ .
The quantum effect on the scale factor becomes large 
as the mass of the point particle becomes large, 
as can be seen from Eq.(\ref{hierarchy}) 
and Fig. \ref{fig:coniclala1cp.eps}.  
In the closed de Sitter universe solution, 
a conical singularity occurs due to the 
presence of the mass $m$,  
and in this case the volume of the universe is small.  
We can summarize our results as follows:  
The size of the quantum effects increases as 
the size of the universe decreases. 
This relation is consistent with our general experience:   
quantum effects are larger for smaller objects. 

As future problems, we are interested in 
determining the quantum effect for the anti-de Sitter universe 
using the dBB interpretation. 
We are also interested in  
the total gravitational energy of a massive point particle 
in a closed universe. 
We plan to study a torus as an example of other topologies  
in the (2+1)-dimensional gravity theory \cite{Hosoya90}. 

Finally, it is worthwhile to note the difference between 
the dBB interpretation and the interpretation by the WKB approach. 
Though the concepts of the two interpretations are very different, 
the basic equations are very similar. 
In the semi-classical region,  
both solutions obtaining by the dBB interpretation and 
by the WKB approach become almost the same, 
and  in addition 
the dBB interpretation requires a  
smooth extension into the classically forbidden region.  
For these reasons, we assert that the dBB interpretation is 
more natural than the WKB approach. 

\vspace{5mm}
\noindent
{\Large \bf Acknowledgements}

\noindent 
We would like to thank Dr. T. Takahashi for useful 
discussions and numerical calculations.   


\end{document}